# Synergistic Effects of Nanoparticle Heating and Amoxicillin on *H. Pylori* Inhibition

Tao Wu[1], Lichen Wang[2], Meiliang Gong[1], Yunjuan Lin[1], Yaping Xu[1], Ling Ye[1], Xiang Yu[2], Jing Liu[1], Jianwei Liu[1], Shuli He[2#], Hao Zeng[3#], Gangshi Wang[1#]

[1] The Second Medical Center, Chinese PLA General Hospital and National Clinical Research Center for Geriatric Diseases, Beijing 100853, China
[2] Department of Physics, Capital Normal University, Beijing 100048, China
[3] Department of Physics, University at Buffalo, SUNY, Buffalo, New York 14260, USA

## Abstract

We report the design and development of a dual-functional magnetic nanoparticle platform for potential treatment of *H. pylori* infection. We show that an ultralow concentration of $Mn_{0.3}Fe_{2.7}O_4$@$SiO_2$ nanoparticles subjected to a moderate AC magnetic field, without bulk heating effect, can deposit heat locally and effectively inhibit *H. pylori* growth and virulence *in vitro*. When coupled with antibiotic amoxicillin, the dual-functional amoxicillin loaded $Mn_{0.3}Fe_{2.7}O_4$@$SiO_2$ further decreases the bacteria survival rate by a factor of 7 and 5, respectively, compared to amoxicillin treatment and nanoparticle heating alone. The synergistic effect can be partially attributed to the heating induced damage to the cell membrane and protective biofilm, which may increase the permeability of antibiotics to bacteria. Our method provides a viable approach to treat *H. pylori* infection, with the potential of reducing side effects and enhancing the efficacy for combating drug resistant strains.

# 1. INTRODUCTION

*Helicobacter pylori* (*H. pylori*) infection is a common chronic infectious disease, affecting approximately 4.4 billion individuals worldwide.[1] *H. pylori* causes chronic active gastritis in all colonized subjects. Furthermore, it can lead to peptic ulcer disease, atrophic gastritis, gastric adenocarcinoma, and mucosa-associated lymphoid tissue lymphoma. [2-4] In 1994, *H. pylori* was classified as a member of first class carcinogen of stomach adenocarcinoma. *H. pylori* uses its flagella to burrow into the mucus lining of the stomach to reach the epithelial cells underneath and adhere to them, [5] where it produces biochemicals such as vacuolating cytotoxin A, which can cause inflammation and potentially carcinogenesis. [6-8] A recent study indicated that *H. pylori* eradication reduced the cumulative incidence of gastric cancer in healthy asymptomatic population, and the effect of *H. pylori* eradication on the prevention of gastric cancer was observed in all ages.[9]

Current consensus recommends traditional bismuth quadruple therapy for 14 days as one of the options for first-line therapy of *H. pylori* infection.[10] Antibiotics such as metronidazole, clarithromycin and amoxicillin are commonly used in the regimens. However, the rapid emergence and dissemination of antibiotic-resistant strains has been documented globally.[8, 11] Resistance of *H pylori* to antibiotics has reached alarming levels worldwide. [12] The reported prevalence of clarithomycin resistance among *H. pylori* was approximately 50% in China, and 40% in Turkey. [13] While the prevalence of metronidazole resistance was as high as 90% in some Asian countries. [13-15] The resistance rate to amoxicillin of *H. pylori* is also rising. [16] Besides, side-effects of antibiotics may make them poorly tolerated by some patients. [17, 18] Therefore, it is imperative to explore new treatment methods that can reduce or even avoid the use of antibiotics.

It has been reported that *H. pylori* survived for less than 1 day at 40 °C or 42 °C [19]. However, human body cannot tolerate a temperature of 42 °C for a time duration of 24 h. On the other hand, thermal energy delivered locally has been employed as a novel method of triggering biological responses for various medical applications.[20-24] Recently, it was reported that bacterial viability was significantly reduced by photothermal effect.[25-27] However, the small penetration depth of light severely limits the clinical potential of photothermal therapy.[28] It is well known that magnetic nanoparticles (NPs) exposed in an alternating current (AC) magnetic field can be used to deposit heat locally for magnetic hyperthermia and for triggering neuronal responses.[29-32] Magnetic hyperthermia has attracted great interests in cancer therapy, due to its ability to penetrate deep into the body and its minimal invasiveness.[33] We therefore hypothesize that magnetic NP local heating may be used for *H. pylori* growth inhibition. We further hypothesize that local heating in conjunction with antibiotics may synergistically enhance *H. pylori* growth inhibition, thus reduce the antibiotic usage and address the problem of antibiotic resistance. Based on the above hypotheses, we designed amoxicillin loaded $Mn_{0.3}Fe_{2.7}O_4$@$SiO_2$ NPs as a dual functional agent for local heating and drug carrier. We report two significant findings based on *in vitro* experiments: (1) Local heating by an ultralow concentration ($\leq 1$ μg/ml) of $Mn_{0.3}Fe_{2.7}O_4$@$SiO_2$ NPs can strongly inhibit the *H. pylori* bacteria growth without bulk heating and (2) using amoxicillin loaded $Mn_{0.3}Fe_{2.7}O_4$@$SiO_2$ NPs as a dual functional platform, combined NP local heating and antibiotics show synergistic enhancement of *H. pylori* growth inhibition, with a bacteria survival rate that is just 1/5 and 1/7 of the values for NP heating and amoxicillin alone. These findings pave the way for clinical treatment of *H. Pylori* infection using our novel approach.

## 2. MATERIALS AND METHODS

*2.1. Materials*

*H. pylori* standard strain NCTC 11637 was kindly provided by the *H. pylori* Strain Pool, Beijing, China. *H. pylori* medium and Broth medium were purchased from Hopebio, China. Cyclohexane and ammonia hydroxide were purchased from Alfa. IGEPAL CO-520 and Tetraethyl orthosilicate (TEOS) were purchased from Sigma-Aldrich Inc., USA. Amoxicillin were purchased from Adamas. *Campylobacter* Agar Base were purchased from OXOID.

## 2.2. Preparation and characterization of $Mn_{0.3}Fe_{2.7}O_4@SiO_2$ NPs

$Mn_{0.3}Fe_{2.7}O_4$ NPs were synthesized by high temperature organic solution phase reaction method. Iron(III) acetylacetonate (2 mmol), manganese(II) acetylacetonate (1 mmol), 1,2-hexadecanediol (10 mmol), oleic acid (3 mmol), oleylamine (3 mmol), and 20 mL benzyl ether was mixed by magnetic stirring. First, the mixture was heated to 120 °C and maintained for 0.5 h under a flow of nitrogen, then heated to 200 °C and kept for 1 h. The temperature was further raised to 300 °C at a heating rate of 10 °C/min, and kept at 300 °C for 1 h. After the reaction, the heat source was removed and the reaction products were cooled down to room temperature. The NPs were precipitated from the solution by adding ethanol and centrifugation. The collected particles were redispersed in hexane for preservation. The NPs were coated with a silica shell by a reverse microemulsion method. 1.15 ml CO-520, 20 ml cyclohexane and 20 mg $Mn_{0.3}Fe_{2.7}O_4$ NPs were mixed in a 100 ml round bottom flask. After mixing for 20min, 0.15 ml ammonium hydroxide (28-30%) was added. Finally, 0.15 ml TEOS were added dropwise, and the solution was stirred for 24 h. The resulting $Mn_{0.3}Fe_{2.7}O_4@SiO_2$ NPs were washed and precipitated by adding ethanol and centrifuged. Finally, the $Mn_{0.3}Fe_{2.7}O_4@SiO_2$ NPs were dispersed in deionized water.

## 2.3. Cytotoxicity of $Mn_{0.3}Fe_{2.7}O_4@SiO_2$ NPs on SGC-7901 cells and H. pylori

Human gastric cancer cell line SGC-7901 cells were seeded in a 96-well plate (3000 cells / 100 μl / well) and cultured overnight. Cells were then incubated with different concentrations of NPs (0 μg/ml, 50 μg/ml, 100 μg/ml, 150 μg/ml, 200 μg/ml, 250 μg/ml, 300 μg/ml) for 24 h or 48 h. After that, CCK-8 solution (10 μl/well) was added to each well and the plate was incubated at 37 °C for 40 min in the dark. The optical density (OD) values of each well were measured by a spectrophotometer at a wavelength of 450 nm to quantify the cell growth in each group.

H. pylori NCTC 11637 were used throughout the experiments. H. pylori was grown in *Campylobacter* agar base containing 5% sheep blood under microaerobic conditions (5% $O_2$, 10% $CO_2$, and 85% $N_2$) at 37°C for 2–3 days. Different concentration of NPs (1 μg/ml, 10 μg/ml, 20 μg/ml, 50 μg/ml, 100 μg/ml, 200 μg/ml) was added into the agar base, then H. pylori was inoculated and cultured for 3 days. Finally, colonies of bacteria were observed. For the morphology and adhesion ability analyses, H. pylori was grown in Broth Medium.

*2.4. Effect of $Mn_{0.3}Fe_{2.7}O_4@SiO_2$ NP heating on H. pylori growth*

H. pylori culturing agar medium containing $Mn_{0.3}Fe_{2.7}O_4@SiO_2$ NPs (1 μg / ml) was prepared. H. pylori in logarithmic growth phase was collected by phosphate buffer solution (PBS), $1.2 \times 10^7$ colony forming unit (CFU) colonies were added to each 3 cm diameter culture dish. After being evenly coated, the H. pylori were cultured in a 37 °C incubator for 12 h. The culture dish was placed in an AC magnetic field coil to heat the culture medium and the temperature of the culture medium was recorded with a fiber-optical thermocouple. The culture dishes after heat treatment were incubated at 37 °C for 72 h. Finally, each group of H. pylori was collected with PBS, and the absorbance of each group was detected by spectrophotometer at a wavelength of 600 nm（optical density measured at 600 nm, $OD_{600}$), which was used to quantify the H. pylori survival rate of each group.

*2.5. Morphology of H. pylori observed by transmission electron microscopy (TEM)*

H. pylori grown in Broth Medium containing $Mn_{0.3}Fe_{2.7}O_4@SiO_2$ NPs (0.4 μg / ml) was placed in an AC magnetic field. The culture medium temperature was raised to 41°C and kept at that temperature for 10 min. It was then incubated at 37 °C for 2 h. The supernatant was discarded after centrifugation for 4 min at a rate of 1500 r/min. The bacterial concentration was adjusted to $10^8$ CFU/ml by using the 3% glutaraldehyde fixation solution to resuspend the H. pylori pellet, and the bacteria were fixed by placing them in a refrigerator at 4 °C. 10 μl of H.

*pylori* bacterial liquid was dropped on the amorphous carbon-coated copper grids and let dry. One drop of 3% phosphotungstic acid dye solution was added for negative staining. TEM (JEOL-2100) was employed to observe the morphological change of the bacteria.

2.6. Evaluation of H. pylori adhesion ability

*H. pylori* grown in Broth Medium was cultured at 37 ℃ in microaerobic environment for 2 h after different treatment. The bacteria were collected (1500 r/min, 4 min) and dissolved in RPMI-1640 medium without antibiotics and serum, to form *H. pylori*-RPMI-1640 solution, and then adjusted to $OD_{600} = 0.1$ for standby. SGC-7901 cells in logarithmic growth phase were seeded in 96-well plate with $1 \times 10^4$ cells/well, and cultured in incubator at 37 °C overnight. The as-treated bacteria liquid were added for co-cultivation for 2 h with the ratio of bacteria: cells = 100: 1. The supernatant was discarded and then the cells were washed with PBS for three times to wash away cell debris and non-adhering *H. pylori*. 100 μl urea reagent was added into each well, and incubated for 2 h at room temperature. The absorbance of each group was measured with a spectrophotometer with a wavelength of 540 nm.

2.7. Preparation of amoxicillin loaded $Mn_{0.3}Fe_{2.7}O_4@SiO_2$ NPs and detection of amoxicillin release

1 ml of amoxicillin standard solution of different concentrations (0 μg/ml, 20 μg/ml, 30 μg/ml, 40 μg/ml, 50 μg/ml, 60 μg/ml) were detected by high performance liquid chromatography (HPLC) (Agilent 1100), and the peak areas of different concentration of amoxicillin was recorded to draw the standard curve of peak area concentration. 1 ml of $Mn_{0.3}Fe_{2.7}O_4@SiO_2$ solution (10 mg/ml) was added into 9 ml amoxicillin solution (1.5 mg/ml), then stirred continuously for 8 h at room temperature. The mixture was placed under a NdFeB magnet for 12 h at room temperature. The NPs were precipitated to the bottom; the supernatant removed and stored under 4 °C. The NPs that were adsorbed to the bottom were dissolved in 1ml double-distilled water. The solutions were then subjected to HPLC to detect the concentration of amoxicillin. The drug-loading of amoxicillin loaded magnetic NPs was therefore deduced. Drug loading content were calculated as follows: rate of drug loading (%) = (total amoxicillin amount – amoxicillin amount in supernatant)/total amoxicillin amount.

*2.8. Detection of surface potential of NPs*

1 mg/ml amoxicillin sodium chloride solution, 1 mg/ml $Mn_{0.3}Fe_{2.7}O_4@SiO_2$ sodium chloride solution and 1 mg/ml amoxicillin loaded $Mn_{0.3}Fe_{2.7}O_4@SiO_2$ sodium chloride solution were prepared, respectively. The surface potential of these samples was measured using a Malvern Zetasizer Nano ZS90 nanoparticle size potential analyzer.

*2.9. Detection of amoxicillin minimum inhibitory concentration (MIC) on H. pylori*

The campylobacter agar mediums with amoxicillin concentration gradient of 20 μg/ml, 10 μg/ml, 5 μg/ml, 2.5 μg/ml, 1.25 μg/ml, 0.625 μg/ml, 0.3125 μg/ml, 0.15625 μg/ml, 0.078125 μg/ml, 0.0390625 μg/ml and 0 μg/ml were prepared respectively, according to the National Committee for Clinical Laboratory Standards (NCCLS). *H. pylori* in logarithmic growth phase was collected by PBS and dispensed to a concentration of $OD_{600} = 1$ (about $3 \times 10^8$ CFU/ ml). 40 μl of the bacteria liquid was added to each 3 cm culture dish, and spread evenly. The culture dishes were placed inertly in microaerobic gas bags and cultured at 37 °C in incubator for 72 h to observe the colony formation of each group.

*2.10. H. pylori growth inhibition by amoxicillin loaded $Mn_{0.3}Fe_{2.7}O_4@SiO_2$ NPs*

The media containing amoxicillin, $Mn_{0.3}Fe_{2.7}O_4@SiO_2$, and amoxicillin loaded $Mn_{0.3}Fe_{2.7}O_4@SiO_2$ NPs were prepared and grouped according to heating/no heating and heating methods. 40 μl of *H. pylori* bacteria solution ($OD_{600} = 1$) was inoculated into a petri dish with diameter of 3 cm, and cultured in incubator at 37 °C for 12 h. Groups B1 and C1 were placed in AC magnetic field, so that the medium temperature reached 41 °C and the temperature was maintained for 10 min; Group D1 was heated in a water bath and maintained at 41 °C for 10 min; Groups A / B2 / C2 / D2 were cultured at 37 °C as control groups. After the treatment described above, the petri dishes were placed in an incubator and cultured at 37 ℃ for 72 h. *H. pylori* in each group was collected by PBS, and the absorbance of each group was detected with a spectrophotometer with wavelength of 600 nm.

*2.11. H. pylori colony forming test*

*H. pylori* agar media containing amoxicillin, $Mn_{0.3}Fe_{2.7}O_4@SiO_2$ and amoxicillin loaded $Mn_{0.3}Fe_{2.7}O_4@SiO_2$ NPs were prepared. $3 \times 10^4$ CFU of *H. pylori* was inoculated into a 3 cm diameter petri dish. The culture dish of group B1 and C1 was heated in an AC magnetic field, and the temperature of the medium was kept at 41 ℃ for 10 min. The D1 medium was placed in a 41 ℃ water bath for 10min. Group A/B2/C2/D2 were unheated control groups and cultured at 37 ℃. After the above treatments, the culture dishes were placed in a microaerobic environment at 37°C for 72 h, and the colony forming units in each group was observed.

*2.12. Statistical analyses*

The experimental data were expressed as mean ± standard deviation (SD). SPSS19.0 (IBM Corp., Armonk, NY, USA) was used to analyze the data. Univariate comparisons between two groups were made using Student's *t* test for normally distributed data, Wilcoxon rank sum for skewed data. Comparisons of measurement data among multiple groups were analyzed by one-way analysis of variance (ANOVA). $P < 0.05$ was indicative of statistical significance.

# 3. EXPERIMENTAL RESULTS

Our approach is shown schematically in Fig. 1. $Mn_{0.3}Fe_{2.7}O_4$ NPs were synthesized by high temperature organic solution phase reaction of metal acetylacetonates and 1,2-hexadecanediol in the presence of oleic acid and oleylamine. [34] Silica coating of NPs was performed using reverse microemulsion method. [35] Amoxicillin was coupled to silica shells of $Mn_{0.3}Fe_{2.7}O_4@SiO_2$ NPs by electrostatic interactions forming amoxicillin loaded

$Mn_{0.3}Fe_{2.7}O_4$@$SiO_2$ NPs. *H. Pylori* growth inhibition were systematically investigated in the presence of AC field heating of magnetic NPs and amoxicillin for the dual functional agent, and compared to that of NP heating and amoxicillin separately.

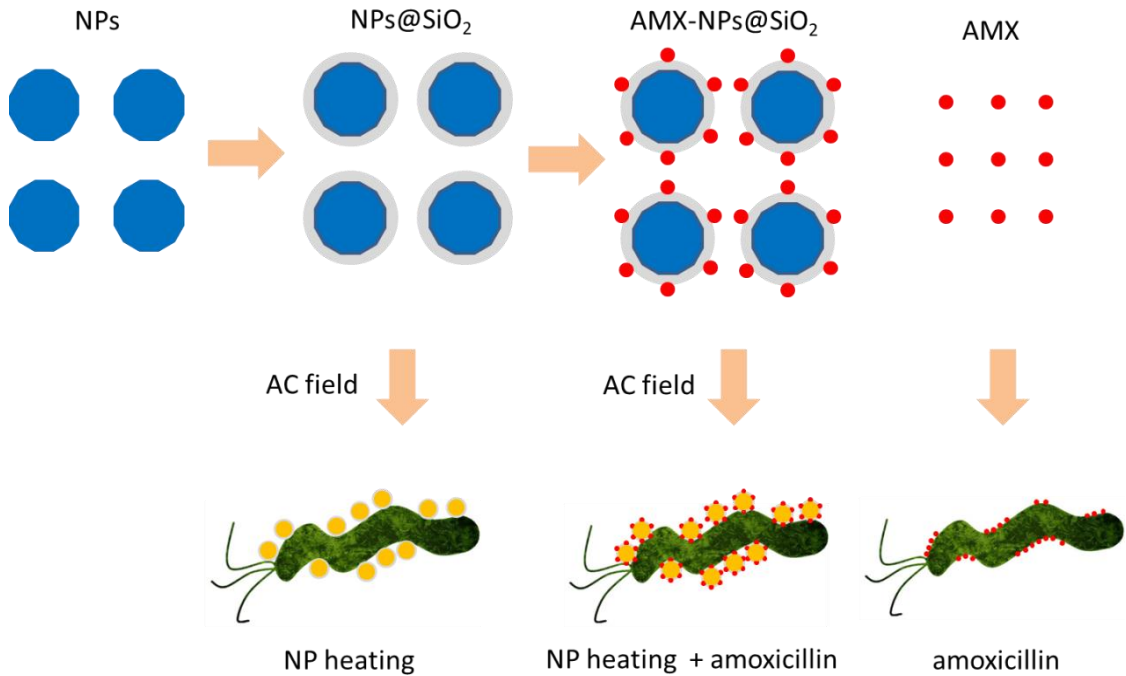

**Figure 1.** A schematic of our approach using amoxicillin loaded $Mn_{0.3}Fe_{2.7}O_4$@$SiO_2$ NPs dual functional agent for *H. Pylori* growth inhibition. The results were compared to using NP heating and amoxicillin separately.

Monodisperse, highly crystalline and high magnetic moment $Mn_{0.3}Fe_{2.7}O_4$ NPs with a diameter of 22 nm were synthesized by a one-pot thermal decomposition method, and used as a model drug carrier system. [34] Our previous study shows that soft ferrite NPs with relatively large size possess high specific loss power (SLP) at moderate AC field amplitudes.[22] However, when the NP sizes are larger than 22 nm, they become ferromagnetic and tend to agglomerate in the solution due to strong magnetostatic interactions. This puts the upper size limit of these ferrite NPs at around 22 nm. Fig. 2 (a) shows a TEM image of monodisperse 22 nm $Mn_{0.3}Fe_{2.7}O_4$ NPs. Electron diffraction pattern shown in the inset of Fig. 2 (a) confirms the spinal structure of the NPs with excellent crystallinity. As-synthesized NPs can be dispersed in hexane in the presence of oleic acid and oleylamine. For bio-applications, the surface of NPs was modified by silica coating to make them water-soluble and biocompatible. Silica has been widely used in biomedical applications, because of its merits of high stability in a wide range of pH [20, 36], ease of functionalization, and biocompatibility. A typical TEM image of $Mn_{0.3}Fe_{2.7}O_4$@$SiO_2$ NPs is shown in Fig. 2 (b). The silica shells were kept thin with a thickness of 4-5 nm, and reaction parameters were controlled so that no free silica particles were present. Aqueous dispersions of $Mn_{0.3}Fe_{2.7}O_4$@$SiO_2$ remain stable without agglomeration for over 2 years. [22]

The magnetic hysteresis loops of $Mn_{0.3}Fe_{2.7}O_4$ NPs were measured at 300 K and 5 K, as shown in Fig. 2 (c). $Mn_{0.3}Fe_{2.7}O_4$ NPs are superparamagnetic at 300 K, while exhibiting ferromagnetism at 5 K (see inset of Fig. 2 (c)). The saturation magnetization $\sigma_S$ of $Mn_{0.3}Fe_{2.7}O_4$ NPs is 96.5 emu/g at 5 K and 85.5 emu/g at 300 K, close to the bulk values and among the highest values for manganese-ferrite NPs. The high saturation magnetization is due to the high crystallinity of the NPs resulting from the polyol synthesis process.[37] The bulk $\sigma_S$ value is 112 emu/g for $MnFe_2O_4$ and 98 emu/g for $Fe_3O_4$ at 0 K,[38] and the bulk $\sigma_S$ value for $Mn_{0.3}Fe_{2.7}O_4$ is expected to be ~ 102 emu/g. We estimate a 5% reduction in $\sigma_S$ for NPs compared to bulk value, which can mainly be attributed to surface spin disorder due to changes in coordination numbers that modifies the exchange interaction.[39] At a diameter of 22 nm, if we assume that Fe + Mn atoms located at up to half a unit cell thick surface layer lose their magnetic moment completely, it would amount to a reduction of 12% in $\sigma_S$. Note however, ferrites have antiferromagnetic super-exchange, and missing nearest neighbor atoms at the surface does not necessarily lead to moment reduction.

The coercivity $H_C$ at 5 K is 205 Oe (16.3 kA/m), from which the magnetic anisotropy energy can be estimated by $M_S H_K \approx M_S H_C$, where $H_K$ is the anisotropy field that can be approximated by $H_C$ at low temperatures. The anisotropy energy value is thus estimated to be $1.0\times10^5$ erg/cm$^3$ ($1.0\times10^4$ J/m$^3$), which is in between the bulk cubic anisotropy constants of MnFe$_2$O$_4$ and Fe$_3$O$_4$. Surface anisotropy may make additional contributions due to surface spin canting, and change the type of anisotropy from cubic to uniaxial.

Other than magnetization, the single most important parameter for evaluating the hyperthermia performance is the $H_C$ of the AC hysteresis. However, most labs are not equipped to measure the magnetic hysteresis loops at the frequency and field amplitude typically used for hyperthermia. On the other hand, DC measurements at room temperature will simply give superparamagnetic hysteresis loops with zero coercivity. As a compromise, one may estimate the anisotropy field $H_K$ at 300 K, assuming that anisotropy decreases as a power of the reduced magnetization:

$$\frac{K_u}{K_{u,0}} \propto \left(\frac{M_S}{M_{S,0}}\right)^n$$

where $n$ is 3 for uniaxial anisotropy [38], and $K_{u,0}$ and $M_{S,0}$ are quantities at 0 K, which can be substituted by the values measured at 5 K without significant error. $H_K$ at 300 K thus estimated is 160 Oe (12.8 kA/m). For AC hysteresis loss to occur, the field amplitude should be larger than $H_K$ or magnetization cannot be reversed, leading to very low SLP.

Heating curves of aqueous dispersions of Mn$_{0.3}$Fe$_{2.7}$O$_4$@SiO$_2$ NPs with a concentration of 1 mg/ml exposed to AC magnetic fields of 380 kHz and field amplitudes ranging from 14 to 33 kA/m were measured by a fiber optical thermometer, as shown in Fig. 2 (d). In the present work, the minimum field amplitude used was 14 kA/m, greater than the estimated $H_K$ of the NPs at 300 K, as SLP is expected to be low for fields below $H_K$. Contribution from pure water under identical conditions was also measured and subtracted as the background. SLP of NPs was extracted using Box-Lucas fitting, since the measurements were performed under non-adiabatic conditions. [40] SLP *vs* magnetic field amplitude was plotted in Fig 2 (e). It can be seen that SLP of Mn$_{0.3}$Fe$_{2.7}$O$_4$/SiO$_2$ NPs increases monotonically with increasing magnetic field amplitude, from ~ 1,200 W/g at H = 14 kA/m to ~ 2,500 W/g at H = 33 kA/m. The intrinsic loss power (ILP), defined as *SLP/H$^2$f*, are 16.4 and 6.0 nHm$^2$/kg, respectively (see Fig. 2 (f)). Note that ILP values are reported here strictly for the purpose of comparison to results from literature. Due to a lack of standards in magnetic hyperthermia measurements, ILP was introduced previously to compare heating performance of NPs measured under different field parameters based on linear response theory, and thus SLP scales linearly with H$^2$ and ILP should be field-independent.[41] This is clearly not the case here, as seen from Fig. 2. Moreover, majority of the magnetic hyperthermia work used field amplitudes exceeding those required by the linear response theory, and thus there is little physical significance of ILP. Our SLP and ILP values are significantly higher than earlier reported results for magnetite and manganese ferrite NPs, [33, 42, 43] suggesting that the Mn$_{0.3}$Fe$_{2.7}$O$_4$@SiO$_2$ NPs possess adequate heating performance, and therefore are good candidates for studying *H. pylori* inhibition by delivering local heat.

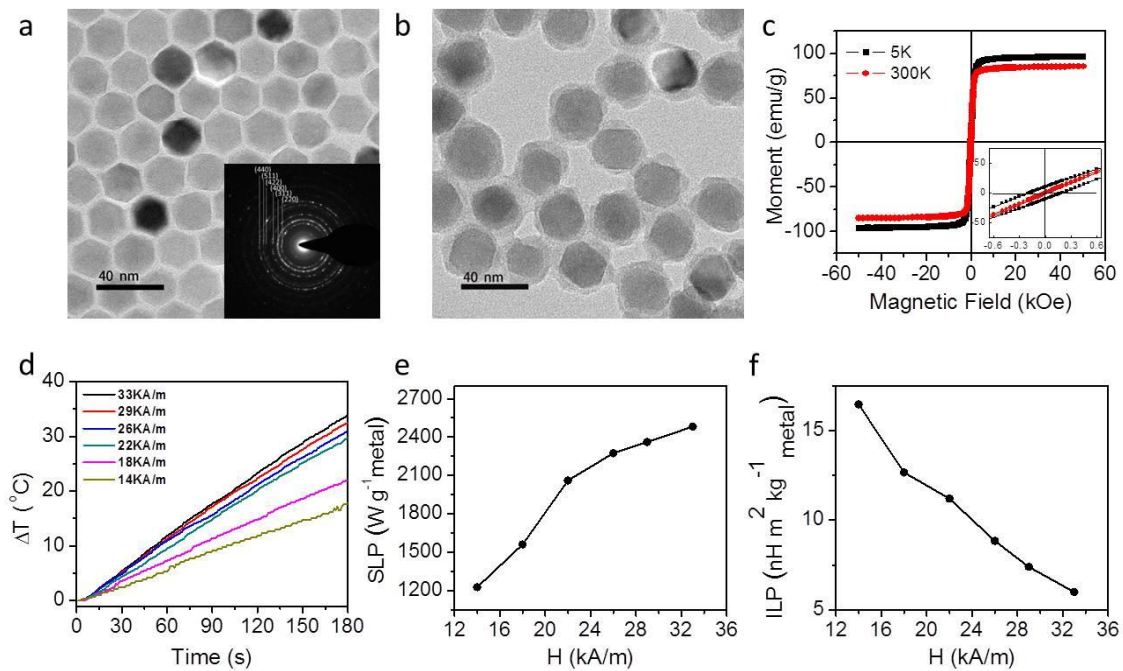

**Figure 2.** (a) A typical TEM image of monodisperse $Mn_{0.3}Fe_{2.7}O_4$ NPs and an electron diffraction pattern of the NPs showing spinel structure; (b) a typical TEM image of $Mn_{0.3}Fe_{2.7}O_4@SiO_2$ NPs; (c) the magnetic hysteresis loops of as-synthesized $Mn_{0.3}Fe_{2.7}O_4$ NPs at temperatures of 5K and 300K, the inset plots the hysteresis loops in a smaller field range, showing ferromagnetic behavior at 5 K; (d) the AC field heating curves measured at 380 kHz and varying field amplitude, for NPs with a concentration of 1 mg/ml; (e) field dependence of SLP for $Mn_{0.3}Fe_{2.7}O_4@SiO_2$ NPs; (f) field dependence of ILP for $Mn_{0.3}Fe_{2.7}O_4@SiO_2$ NPs

To study the effects of NP local heating on the inhibition of *H. pylori* growth *in vitro*, *H. pylori* culturing agar medium containing 1 µg/ml $Mn_{0.3}Fe_{2.7}O_4@SiO_2$ NPs were subjected to an AC magnetic field of 380 kHz and varying amplitude for 15 minutes. This chosen NP concentration is three orders of magnitude lower than what is needed for bulk solution heating (Fig. 2 (e)), and thus the heat energy deposited by NPs is insignificant to raise the agar medium background temperature. The AC magnetic field itself does lead to background heating of the campylobacter agar medium. For example, the top panel in Fig. 3 (a) shows the measured temperature of the agar medium subjected to an AC field of 17.5 kA/m for 15 min. After 7 min heating, the background temperature ($T_b$) was stabilized at 41 °C. Adding NPs with a concentration of 1 µg/ml leads to almost no change in the background heating profile, as seen from the bottom panel in Fig 3 (a). When the agar medium was heated at 42 °C by an AC field for a time duration of 15 min, this background heating at $T_b$ did not alter the viability of *H. Pylori* (Fig. 3 (b)), but can serve as a measure of the local heat energy deposited by NPs on the bacteria (which cannot be measured directly due to limited spatial resolution of the thermocouple), as both would increase with the field amplitude proportionally. $OD_{600}$ value of each group was used to quantify the bacterial growth inhibition rate.

The growth inhibition rate of *H. pylori* cultured on NPs-containing agar medium increases monotonically with increasing $T_b$ from 37 °C (14 kA/m) to 42 °C (18 kA/m) for a fixed heating time of 15 min, as shown in Fig. 3 (c). It is seen that with increasing AC field amplitude so that $T_b$ in the medium was raised to 41 °C, the inhibition rate of *H. pylori* growth was 75%. This value further increases to 96% when $T_b$ was increased to 42 °C. The time-dependent *H. pylori* inhibition at $T_b$ = 42 °C was investigated, as shown in Fig. 3 (d). With the extension of heating time, the growth inhibition rate of *H. pylori* increases monotonically from 18% for the heating time of 1 min to 96% for 15 min.

Cytotoxicity of NPs to SGC-7901 cells and *H. Pylori* was investigated in order to rule out the direct inhibitory effect of $Mn_{0.3}Fe_{2.7}O_4@SiO_2$ NPs on both mammalian and bacterial cell growth. Compared to the group without NPs, the group containing 100 μg/ml $Mn_{0.3}Fe_{2.7}O_4@SiO_2$ NPs showed no significant difference in cell viability at the incubation time of up to 48 h (Fig. 4). The inhibition of *H. pylori* growth by AC field background heating of the agar medium without NPs and by water bath at 42 °C, for a time duration of 15 min, was also measured for comparison. No significant growth inhibition was observed in either case (Fig. 3 (b)). This suggests that the *H. pylori* inhibition is not due to the global heating of the medium, but due to the local heat energy deposited on the bacteria surfaces immediately adjacent to the NPs. Previous studies by our group and others suggest that during AC field heating, the local temperatures at the cell membranes immediately adjacent to the NPs are significantly higher than the environment temperature measured by a thermometer, albeit the low spatial resolution of the fiber-optical thermocouple prevents direct measurement of such local temperatures. [21, 44] Thus AC field heating of magnetic NPs is advantageous as it can deliver targeted heating without raising the environment temperature significantly and causing collateral damages to healthy cells or tissues.

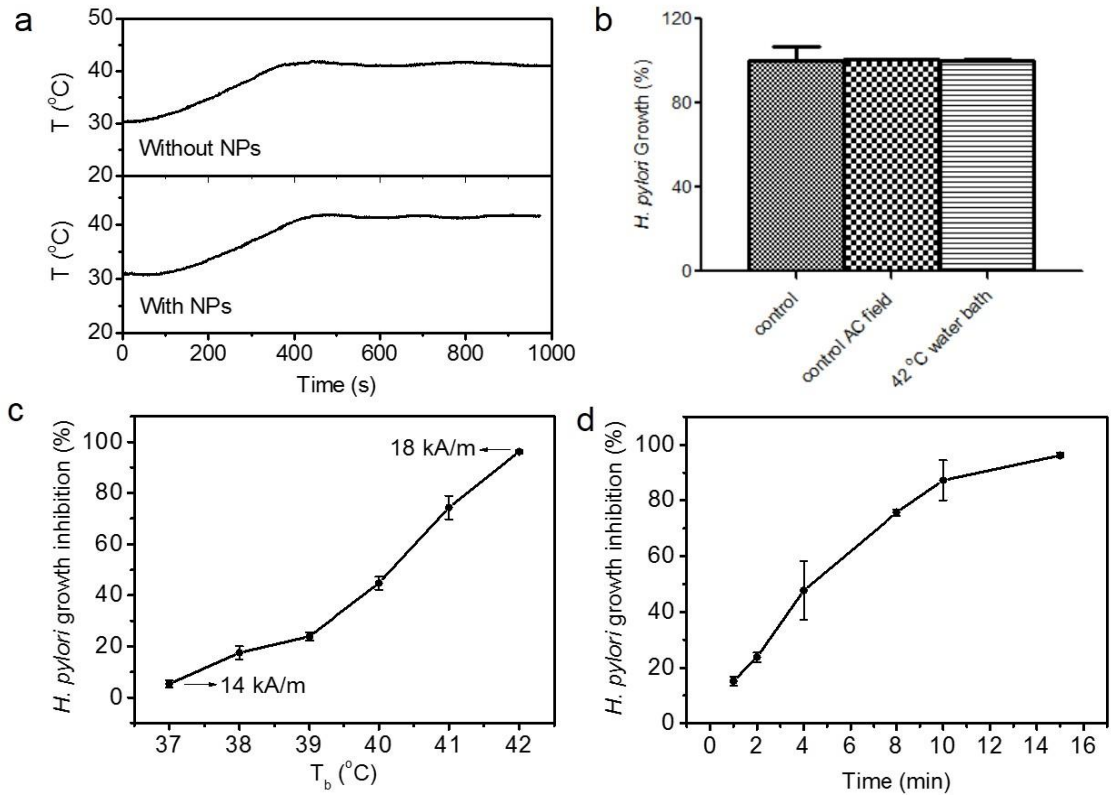

**Figure 3.** (a) Temperature *vs* time curve of *H. pylori* culturing agar medium exposed in AC field of 380 kHz, 17.5 kA/m. Top panel: without NPs; bottom panel: with addition of 1 μg/ml NPs. The medium temperature $T_b$ was stabilized at 41 °C ± 0.5 °C; (b) *H. Pylori* growth under the AC field without NPs and in water bath heated at 42 °C for 15 min. No significant growth inhibition was observed in each group; (c and d) Effect of $Mn_{0.3}Fe_{2.7}O_4@SiO_2$ heating on *H. pylori* growth inhibition: Percentage of *H. Pylori* growth inhibition (c) as a function of $T_b$, t =15 min under the field of 380 kHz; (d) as a function of time, $T_b$ = 42 °C under the field of 380 kHz, 18 kA/m.

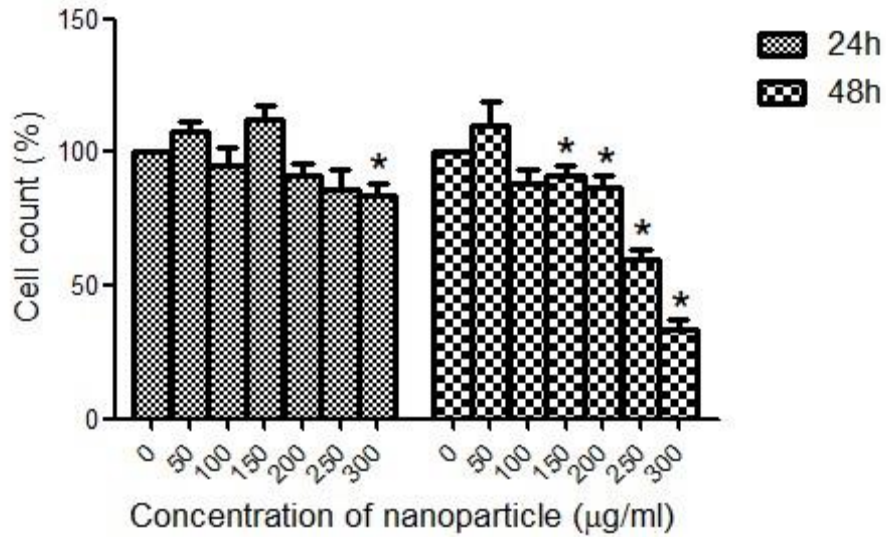

**Figure 4.** Cytotoxicity of $Mn_{0.3}Fe_{2.7}O_4@SiO_2$ NPs to SGC-7901 cells. * P<0.05, compared with 0 µg/ml group (control group).

The above experiments confirm that NP local heating is effective in *H. pylori* inhibition at an ultralow NP concentration of 1 µg/ml incapable of inducing bulk heating. To elucidate the mechanisms of *H. pylori* growth inhibition, TEM images of the bacteria before and after NP heating ($T_b$= 41 °C for 15 min) was done to investigate possible changes in bacteria morphology. According to Fig. 3 (c), the bacteria growth inhibition was 75% under this treatment condition. It was found that the morphology of the *H. pylori* after NP heating changed dramatically, as seen in Fig. 5. Untreated *H. pylori* typically appears to be a unipolar, blunt-ended rod with a length of about a few µm and a flagellum with sheath on one end, and the density of the whole cell body was uniform, as shown in Fig. 5 (a) and (b). In the close-up view of a single bacterium, a scrub-like biofilm was observed surrounding the cell after negative staining (Fig. 5 (b)). The biofilm is a matrix containing extracellular polymeric substances to embed bacteria, [9] which plays a main role in the antibiotic resistance. After NP heating, however, the morphology of the *H. pylori* changed dramatically, as seen from Fig. 5 (c) and 5(d). The flagellum and the scrub-like biofilm disappeared. The cell body became swollen with cell membrane damaged. The density of the cell decreased non-uniformly, suggesting that parts of the interior of the cell body became hollow. A high concentration of magnetic NPs was observed to accumulate on the surface of the bacteria. In the more extreme cases as shown in Fig. 5 (d), the bacteria changed shape completely from rod-like to being bent like a "V" shape. Such dramatic changes in morphology indicate direct growth inhibition upon NP heating, consistent with the inhibition rate of 75% at $T_b$= 41 °C shown in Fig. 3 (c). Based on the TEM results, we propose the following mechanisms of *H. pylori* growth inhibition by NP heating: 1. The local heating causes cell membrane rupture, releasing its content, and leads to direct bacteria death; and 2. As the flagellum of normal *H. pylori* can make it swim through gastric mucus and bind to the gastric epithelial cells through the secretion of adhesion, the loss of flagella will cause the bacteria to lose its mobility and adhesion ability.

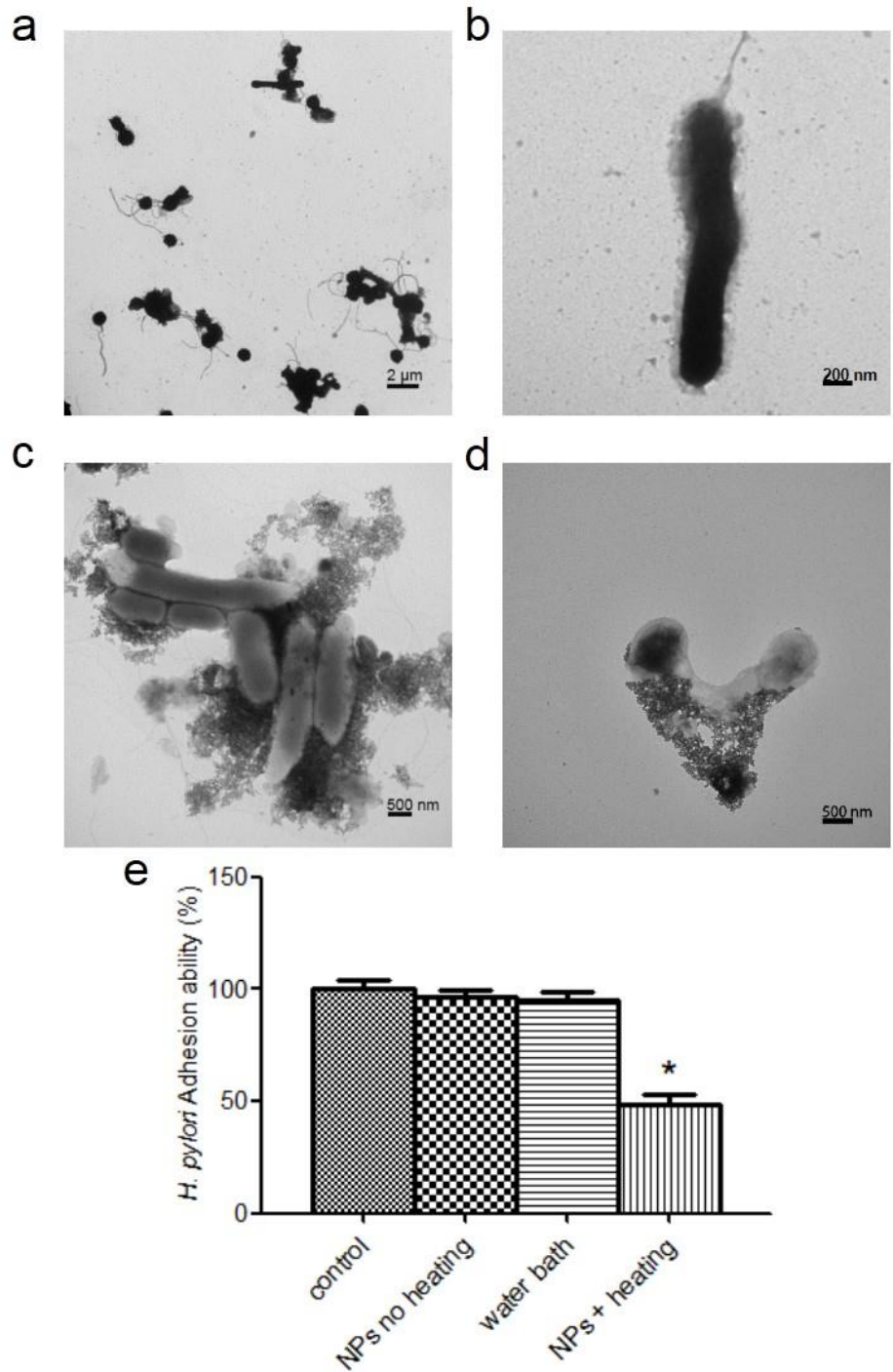

**Figure 5.** Effects of NP heating on *H. pylori* NCTC11637 morphology based on transmission electron microscopy. (a) and (b) Untreated normal *H. pylori*; (c) and (d) NP heating treated *H. pylori*, showing bacteria with lost flagellum and bio-films; (e) changes of *H. pylori* adhesion ability after NP heating treatment. Compared with control groups, the group with $Mn_{0.3}Fe_{2.7}O_4@SiO_2$ NP heating showed significant inhibition of the adhesion ability. * *P* <

0.05, vs. the control group (n = 5).

The changes of the *H. pylori* adhesion ability after NP heating were measured. Compared with the control group, the NPs heating group, which was heated in AC magnetic field at $T_b$ = 41 °C for 10 min, showed an adhesion inhibitory rate of 51.78%, while either water bath group or NP without heating group exhibited no significant statistical difference as seen from Fig. 5 (e). These results confirmed one major mechanism of NP heating induced *H. pylori* growth inhibition: the localized heating leads to morphological changes of the bacteria. Specifically, it causes the bacteria to lose its flagellum, which prevents *H. pylori* from adhering to cells.

The NP heating induced bacteria growth inhibition, together with the observed change in bacteria morphology and adhesion ability is interesting. It seems to be inconsistent with what was predicted from classical heat diffusion model. Using this model, it has been suggested that at a concentration of 1 mg/ml, typical magnetic NPs with a SLP of a few 100 W/g are incapable of reaching the therapeutic temperature of 42 °C for a tumor size smaller than 3 mm. [45] However, our effective NP concentration was three orders of magnitude lower. The sizes of the cultured bacteria clusters, on the other hand, were 1 mm or smaller. Recently, however, more experimental evidence emerged to show that local temperatures in the vicinity of cell surfaces may be dramatically different from the global temperature measured by a thermocouple. [46, 47] We propose two scenarios that might explain this apparent discrepancy: 1. The local concentration of NPs in the vicinity of the bacteria surface is much higher than the nominal one, as evidenced in Fig. 5. The collective behavior of these NPs can heat the bacteria surface to temperatures much higher than those measured by a fiber-optical thermocouple, which does not have the spatial resolution to resolve the local temperature. These local "hot spots" may be the ones making dominant contributions to thermal energy deposition and cell damage; 2. The ordered structure of cell membranes may lead to quasi-ballistic thermal transport, when the size of the heat generating region is smaller than the phonon mean free path. This may lead to a much higher temperature rise and a sharp temperature gradient at the cell surface.[48] Future studies focusing on non-uniform distribution of NPs in biological systems and reliable characterization of local temperature at the nanoscale are needed to clarify the origin of the observed local heating. Such studies are of great significance as nanoscale local heating may enable treatment of small tumors and metastasis cancer cells.

Although the direct growth inhibition rate was at 75% for $T_b$= 41 °C for 15 min, the remaining survival bacteria may still be affected by NP heating. From the TEM studies, a universal feature exhibited by almost all bacteria was that the biofilms were destroyed by NP heating. A recent study reported that local hyperthermia induced by iron oxide nanoparticles can disperse biofilms of *Pseudomonas aeruginosa*, and increase its susceptibility towards antimicrobials. [49] We postulate that if antibiotics were administered simultaneously with NP heating to *H. pylori*, there would be a synergistic effect on bacterial inhibition. As amoxicillin is presently one of the most commonly recommended antibiotics used for *H. pylori* eradication during clinical practices, it would be interesting to study the combined effects of amoxicillin and NP heating. To test this aspect, we designed an amoxicillin loaded $Mn_{0.3}Fe_{2.7}O_4@SiO_2$ dual functional NP platform, relying on electrostatic coupling between the antibiotic and NPs. It is known that NPs with a silica shell possess high negative surface charges. [50] On the other hand, amoxicillin molecules were detected to have a zeta potential level showing small positive charges. [51] We postulate that $Mn_{0.3}Fe_{2.7}O_4@SiO_2$ NPs and amoxicillin can be coupled directly by electrostatic interactions to form amoxicillin loaded $Mn_{0.3}Fe_{2.7}O_4@SiO_2$. The zeta potential was measured to confirm the coupling. It was found that the zeta potential of amoxicillin was 7.022 mV and that of $Mn_{0.3}Fe_{2.7}O_4@SiO_2$ NPs was -46.02 mV, as shown in Fig. 6(a) and (b), respectively. After coupling, the zeta potential value of amoxicillin-$Mn_{0.3}Fe_{2.7}O_4@SiO_2$ was measured to be -40.02 mV (Fig. 6 (c)). This shows that the presence of positively charged amoxicillin reduces the negative surface charge of $Mn_{0.3}Fe_{2.7}O_4@SiO_2$ NPs, suggesting successful binding of the antibiotic to NPs.

The MIC of amoxicillin to *H. pylori*, *i.e.* the lowest concentration of amoxicillin which prevents visible growth of the bacteria, was found to be 0.078 μg/ml. To study potential synergistic effects of NP heating and amoxicillin delivery, 0.04 μg/ml amoxicillin, about a half of the MIC was chosen as the antibiotic concentration used in this experiment. As the drug loading ratio of amoxicillin loaded $Mn_{0.3}Fe_{2.7}O_4@SiO_2$ was measured to be 10.4%, the concentration of $Mn_{0.3}Fe_{2.7}O_4@SiO_2$ NPs used was 0.4 μg/ml. Note that the NP concentration was below the values of 1 μg/ml used above for testing heating effect alone (Fig. 3). *H. pylori* was treated by different methods summarized in Table 1 for comparing their effectiveness of growth inhibition. When AC field heating was present, the field amplitude was controlled so that the background temperature $T_b$ was kept at 41 °C for 10 min (Fig. 3 (a)). The bacterial colony forming units of *H. pylori* were shown in Fig. 6(d), revealing qualitative difference after different treatments. The quantitative bacteria growth inhibition measured by $OD_{600}$ was further plotted in Fig. 6(e). The control groups, namely $Mn_{0.3}Fe_{2.7}O_4@SiO_2$ NPs without heating (group C2) and the group treated at 41 °C in water bath (group D1) had no effect on *H. pylori* growth inhibition. Compared with the control groups, amoxicillin treatment group (group A), $Mn_{0.3}Fe_{2.7}O_4@SiO_2$ NP heat treatment group (group C1) and amoxicillin loaded $Mn_{0.3}Fe_{2.7}O_4@SiO_2$ NP heat treatment group (group B1) all showed significant growth inhibition. The effect of group C1 is significantly better than that of group A. As can be seen from Fig. 6(e), amoxicillin treatment alone (group A) results in cell survival rate of 43.7%, while NP heating alone (group C1) leads to cell survival of 23.7%, which is nearly 50% reduction compared to group A. The inhibition effect of the amoxicillin loaded $Mn_{0.3}Fe_{2.7}O_4@SiO_2$ heat treatment group (group B1) is the highest, with the cell survival being only 6.15% of that of the blank control group. From these data, it can be seen that the NP heating combined with amoxicillin treatment greatly enhances *H. pylori* growth inhibition, decreasing the bacteria survival rate by a factor of 7 and 5, respectively, compared to amoxicillin treatment and NP heating alone.

**Table 1**. Summary of groups with different treatment methods

| Group | AC magnetic field | Reagents/treatment added | Final concentration of reagents added | Inhibitory rate(%) |
|---|---|---|---|---|
| A | No | Amoxicillin | 0.04μg/ml | 56.34 |
| $B_1$ | Yes | Amoxicillin-$Mn_{0.3}Fe_{2.7}O_4/SiO_2$ | 0.4μg/ml（Amoxicillin: 0.04μg/ml） | 93.86 |
| $B_2$ | No | Amoxicillin-$Mn_{0.3}Fe_{2.7}O_4/SiO_2$ | 0.4μg/ml（Amoxicillin: 0.04μg/ml） | 41.47 |
| $C_1$ | Yes | $Mn_{0.3}Fe_{2.7}O_4/SiO_2$ | 0.4μg/ml | 76.26 |
| $C_2$ | No | $Mn_{0.3}Fe_{2.7}O_4/SiO_2$ | 0.4μg/ml | 11.87 |
| $D_1$ | No | Water bath | NA | 18.28 |
| $D_2$ | No | No | NA | 0 |

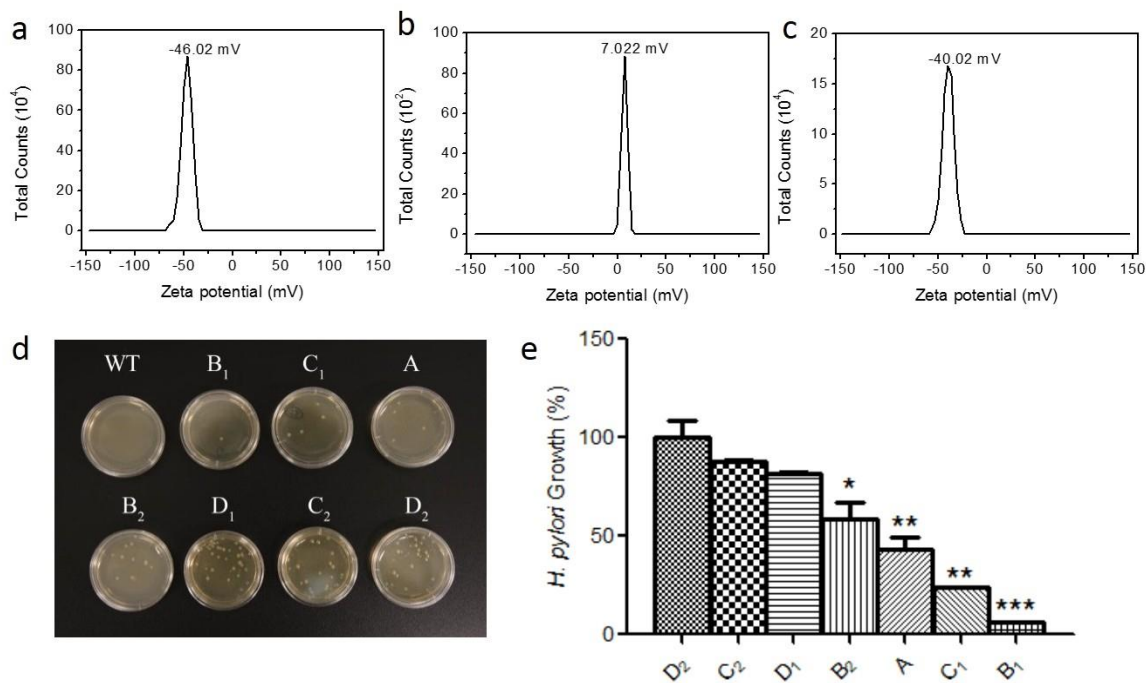

**Figure 6.** Zeta potential of (a) $Mn_{0.3}Fe_{2.7}O_4@SiO_2$ NPs; (b) amoxicillin and (c) amoxicillin loaded $Mn_{0.3}Fe_{2.7}O_4@SiO_2$ NPs; (d) Bacterial colony forming units of *H. pylori* in different treatment groups, D2 served as positive control, WT was negative control; (e) *H pylori* growth after different treatment. Compared with control group, $Mn_{0.3}Fe_{2.7}O_4@SiO_2$ NP with heat treatment group, amoxicillin loaded $Mn_{0.3}Fe_{2.7}O_4@SiO_2$ NP with/without heat treatment group and amoxicillin group all showed significant growth inhibition. The inhibitory effect of the amoxicillin loaded $Mn_{0.3}Fe_{2.7}O_4@SiO_2$ heat treatment group is the highest. *, $P < 0.05$, **, $P < 0.01$, ***, $P < 0.001$, vs. the control group (n = 3).

## 3. DISCUSSIONS

The great enhancement in *H. Pylori* growth and virulence inhibition by amoxicillin loaded $Mn_{0.3}Fe_{2.7}O_4@SiO_2$ NP heat treatment is exciting. The inhibition rate of combined treatment is significantly higher than that of the simple addition of the two separate treatments, confirming synergistic effects. The synergy can be understood as follows: the NP local heating causes dramatic changes in the morphology of the bacteria. In particular, the heating can open up or destroy the biofilm, as seen from Fig. 5(c) and (d). It is known that the biofilm of *H. pylori* is a protective strategy for the bacteria. In addition to helping combating the host immune system and environmental stresses, it can also protect the bacteria from antimicrobial drug treatment, resulting in drug resistance. The loss of the biofilm by NP heating can make *H. Pylori* more susceptible to antimicrobial amoxicillin, and thus enhance its efficacy. The reverse may also be true, that is the low dosage of amoxicillin below MIC may make *H. Pylori* more sensitive to NP heating, thus increase the direct bacteria apoptosis by NP heating.

The synergistic effects of antibiotics and NP heating may greatly reduce the dosage of antibiotics required in the treatment of *H. pylori* infection. The approach is particularly useful for the treatment of drug resistant strains. Because of high bacterial load [52] and the gastric niche, [53] where *H. pylori* could be protected from both antibiotic action and immune response, the drugs has to remain in high concentration for prolonged period in the stomach to display its therapeutic effectiveness. The DC magnetic field gradient can in principle be applied to localize the NPs in the upper-abdominal area by *e.g.* wearing a permanent magnet patch, and to serve as a part of the gastroretentive drug delivery system. By removing the DC field after the treatment, the NPs can be readily excreted through the gastrointestinal tract. As opposed to systemic delivery which inevitably leads to accumulation of NPs in organs, our approach will have less concern of the potential toxicity of the NPs. We therefore envision that this approach can be readily adopted for clinical applications.

## 4. CONCLUSION

In this study, we designed an amoxicillin loaded $Mn_{0.3}Fe_{2.7}O_4@SiO_2$ dual functional NP platform and demonstrated its antibacterial effect under AC magnetic field heating. It is found that the *H. pylori* growth is significantly suppressed by a factor of 7 and 5, respectively, compared to using amoxicillin or NP heating separately, indicating the synergistic effect of NP local heating and drug release. NP local heating can not only lead to direct cell disruption and loss of adhesion ability, but also enhance the susceptibility of *H. pylori* to amoxicillin due to destruction of cell membrane and microbial biofilm. Our approach provides a new pathway for fighting *H. pylori* infection, especially for drug resistant ones. Furthermore, the coupling of antibiotics to magnetic NPs and utilizing the synergistic effects of NP local heating and drug release is a general concept, and can be adopted to fight other infectious diseases with enhanced efficacy.

**AUTHOR INFORMATION**

Corresponding Authors

*E-mail: shulihe@cnu.edu.cn

*E-mail: haozeng@buffalo.edu

*E-mail: wanggangshi@hotmail.com

**ORCID**

Hao Zeng: 0000-0002-6692-6725

Gangshi Wang: 0000-0003-3831-6871

Shuli He: 0000-0002-0333-0553

**ACKNOWLEDGMENTS**

This work was supported by National Science Foundation of China (Grant No. 51471186, 51571146, 51771124).